\documentclass[twocolumn,showpacs]{revtex4}
\usepackage[dvips]{graphicx}
\usepackage{amsmath,amssymb,bm}

\newcommand{\cE}{\mathcal{E}}
\newcommand{\bp}{{\bm p}}
\newcommand{\br}{{\bm r}}

\begin{document}
\title{Periodic-orbit bifurcations as the origin of nuclear deformations}
\author{Ken-ichiro Arita}
\affiliation{Department of Physics, Nagoya Institute of Technology,
Nagoya 466-8555, Japan}
\date{July 15, 2005}
\begin{abstract}
Semiclassical analysis of shell structures in realistic nuclear
potentials are presented using periodic-orbit theory.  We adopted
$r^\alpha$ potential model and examined classical-quantum
correspondence using Fourier transformation technique.  Spin-orbit
coupling is also taken into account in the model Hamiltonian.  Gross
shell structure for a certain combination of surface diffuseness and
spin-orbit parameters are investigated and its relation to pseudospin
symmetry is discussed.  Analysis of superdeformed shell structure in
realistic model is also presented.
\end{abstract}
\maketitle

\section{Shell structure and periodic orbits}

Nuclear deformations are intimately related with shell structures in
single-particle energy spectra of deformed Hamiltonian.  Using the
semiclassical theory, quantum level density $g(E)$ can be represented
in terms of classical periodic orbits, and one obtains the trace
formula\cite{Gutzwiller,Brack}
\begin{equation}
g(E)\sim g_0(E)+\sum_\beta A_\beta(E)\cos\left[\frac{S_\beta(E)}{\hbar}
 -\nu_\beta\right].
\label{eq:trace1}
\end{equation}
$g_0(E)$ is average part of the level density, and the oscillating
part is expressed as the sum over all periodic orbits $\beta$ in
corresponding classical Hamiltonian system.
$S_\beta=\oint_\beta\bp\cdot d\br$ is action integral along the
orbit $\beta$, and $\nu_\beta$ is Maslov phase determined by the
number of conjugate points along the orbit.

In the trace formula, each periodic orbit contribution is an
oscillatory function of energy since action $S_\beta(E)$ is an
increasing function of $E$.  The energy scale of this oscillation
is given by $\delta E\approx 2\pi\hbar/T_\beta$.
%\begin{equation}
%\delta S_\beta=\pp{S_\beta}{E}\delta E=T_\beta \delta E
%\approx 2\pi\hbar, \quad
%\delta E \approx \frac{2\pi\hbar}{T_\beta}
%\end{equation}
From this relation, one can see that short periodic orbits (having
small $T_\beta$) is associated with gross structure (having large
$\delta E$).

The amplitude factor $A_\beta$ in the trace formula (\ref{eq:trace1})
has significant dependence on the stability of the orbit.  In standard
stationary-phase approximation, it is proportional to the stability
factor;
\begin{equation}
A_\beta \propto \frac{1}{\sqrt{|\det(M_\beta-I)|}},
\label{eq:stability}
\end{equation}
where $M_\beta$ represents symmetry-reduced monodromy
matrix\cite{ExtendedGT}, which characterizes linear stability of the
periodic orbit.  Varying external parameters in the Hamiltonian, each
periodic orbit continuously changes its properties, and one of the
eigenvalues of $M_\beta$ might eventually approaches 1.  At this
point, a continuous family of quasi-periodic orbits appear in
neighborhood of the periodic orbit $\beta$ in direction to eigenvector
belonging to the above unit eigenvalue of $M_\beta$.  It usually
accompany the appearance of new periodic orbit from that local family
(or, inversely, disappearance of another periodic orbit into the
family), namely bifurcation of periodic orbit occur at this point.
One should also note that this periodic-orbit bifurcation is related
with the restoration of local dynamical symmetry in neighborhood of
the periodic orbit.  Since these quasi-periodic orbits make coherent
contribution in the periodic orbit sum, we can expect a significant
enhancement of shell effect in quantum level density.  Actually, our
previous works\cite{AM,ASM} clearly show that the bifurcations of
short periodic orbits play essential role in emergence of gross shell
structures.  The divergence of stability factor (\ref{eq:stability})
at bifurcation point is due to the breakdown of standard
stationary-phase method, and can be remedied by using suitable
approximation for trace integral such as uniform
approximations\cite{Ozorio,Sieber} and improved stationary phase
method\cite{Magner}.

\section{The $\bm{r^\alpha}$ potential model}

For nuclei and metallic clusters, central part of the mean field
is well described by Woods-Saxon potential.  For stable nuclei,
Woods-Saxon potential can be further approximated by much simpler
potential with $r^\alpha$ radial dependence\cite{Arita2004}
\begin{equation}
V_{\mathrm{WS}}=-\frac{W}{1+\exp((r-R)/a)}
\approx -W+U\cdot(r/R)^\alpha.
\end{equation}
Here, the parameter $\alpha$ controls the surface diffuseness.
With typical value of Woods-Saxon parameters for nuclei, we found that
the quantum energy spectra of $r^\alpha$ model nicely reproduce those
of Woods-Saxon model up to about $-10$~MeV.

Shifting the zero-point energy, we write the $r^\alpha$ model
Hamiltonian as
\begin{equation}
H=\frac{p^2}{2M}+U\cdot\left(\frac{r}{Rf(\theta,\phi)}\right)^\alpha.
\end{equation}
The advantage of the $r^\alpha$ model is its scaling property.  Since
equations of motion (EOM) are invariant under scale transformation
\begin{equation}
\bm{r}\to c^{\frac1\alpha}\bm{r}, \quad
\bm{p}\to c^{\frac12}\bm{p}, \quad
t\to c^{\frac1\alpha-\frac12}t, \quad
\mbox{(as $E\to cE$)}
\label{eq:scale}
\end{equation}
one has the same set of periodic orbits in all energy surfaces.
The action integral has simple energy dependence
\begin{equation}
S_\beta(E)=\oint_\beta \bm{p}\cdot d\bm{r}
\propto \sqrt{ME}\cdot R(E/U)^{1/\alpha}=\hbar\cE,
\end{equation}
where we define \textit{scaled} energy $\cE$ as
\begin{equation}
\cE=\frac{\sqrt{MUR^2}}{\hbar}
\left(\frac{E}{U}\right)^{1/\alpha+1/2}.
\label{eq:scaledE}
\end{equation}
We also define \textit{scaled} period by
$\tau_\beta=S_\beta(E)/\hbar\cE$, which is a scale-invariant
parameter proper to the periodic orbit $\beta$.
Then we obtain semiclassical level density with scaled energy
\begin{equation}
g(\cE)=g(E)\frac{dE}{d\cE}=g_0(\cE)+\sum_\beta A_\beta(\cE)
{\cos\left(\cE\tau_\beta-\nu_\beta\right)}.
\label{eq:traceformula}
\end{equation}
This simple form is convenient for Fourier analysis.
The Fourier transform of the trace formula
\begin{equation}
F^{\rm cl}(\tau)=\int d\cE\,e^{i\tau\cE}g(\cE)
=F_0(\tau)+\pi\sum_\beta
 e^{i\nu_\beta}\hat{A}_\beta \delta(\tau-\tau_\beta)
\end{equation}
is a function having peaks at scaled period $\tau_\beta$ of classical
periodic orbits.  It suggests that we can extract information on
classical periodic orbits from the Fourier transform of quantum
level density
\begin{equation}
{F^{\rm qm}(\tau)=\int d\cE\,e^{i\tau\cE}g(\cE)}
=\sum_i e^{i\tau\cE_i},
\end{equation}
where $\cE_i$ is related with quantum energy level $E_i$ via
Eq.~(\ref{eq:scaledE}).

\begin{figure}[p]
\begin{center}
\includegraphics[width=.8\columnwidth,bb=50 50 554 712,clip]{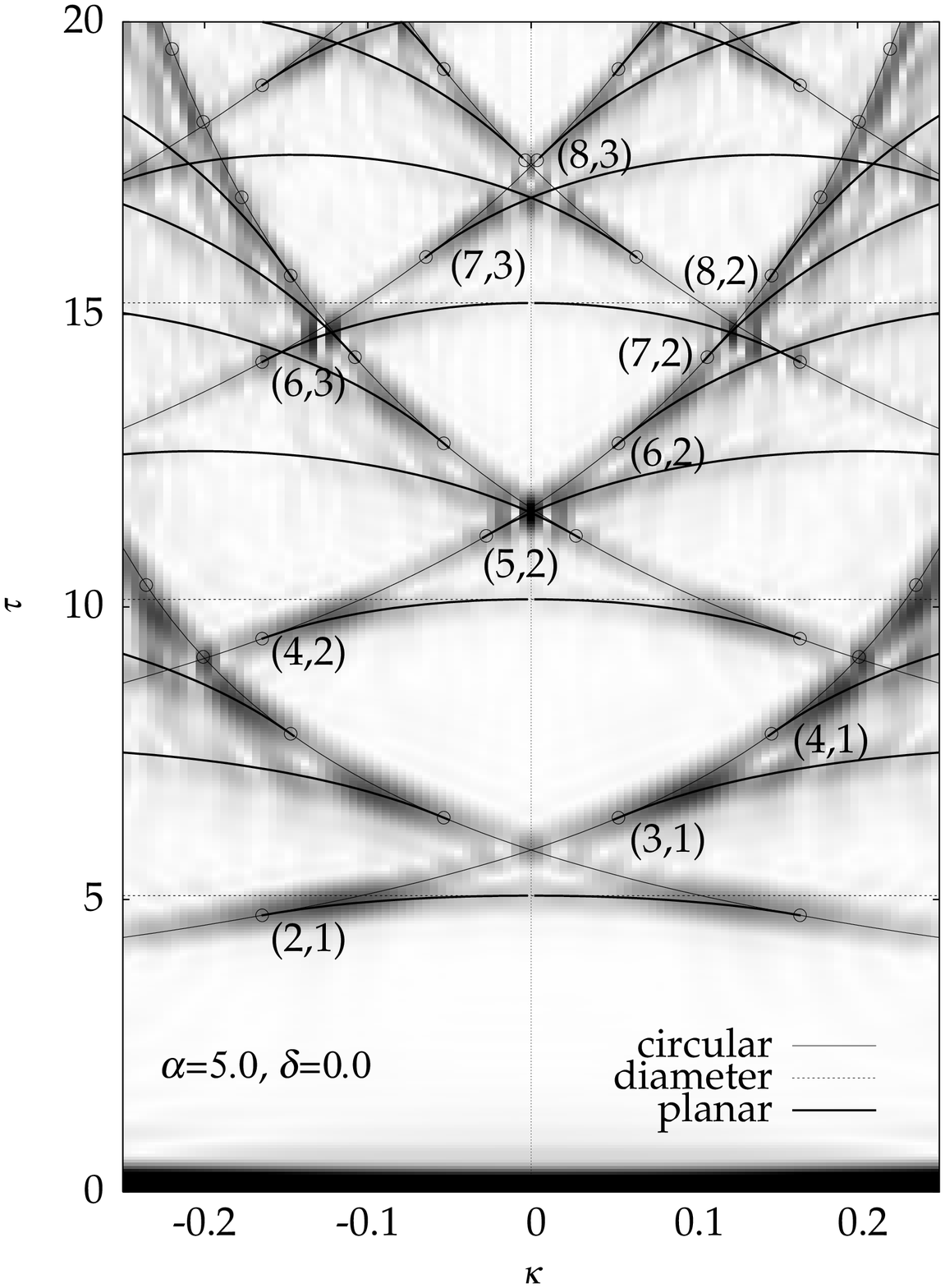}
\end{center}
\caption{\label{fig:fourier}
Scaled periods $\tau_\beta$ of classical periodic orbits plotted as
functions of spin-orbit parameter $\kappa$.  Open circles represent
the bifurcation points.  The background gray-scale image show the
Fourier transform of quantum scaled-energy level density; the Fourier
amplitude $|F(\tau;\kappa)|$ have peaks in dark regions.}

\begin{center}
\includegraphics[width=.85\columnwidth]{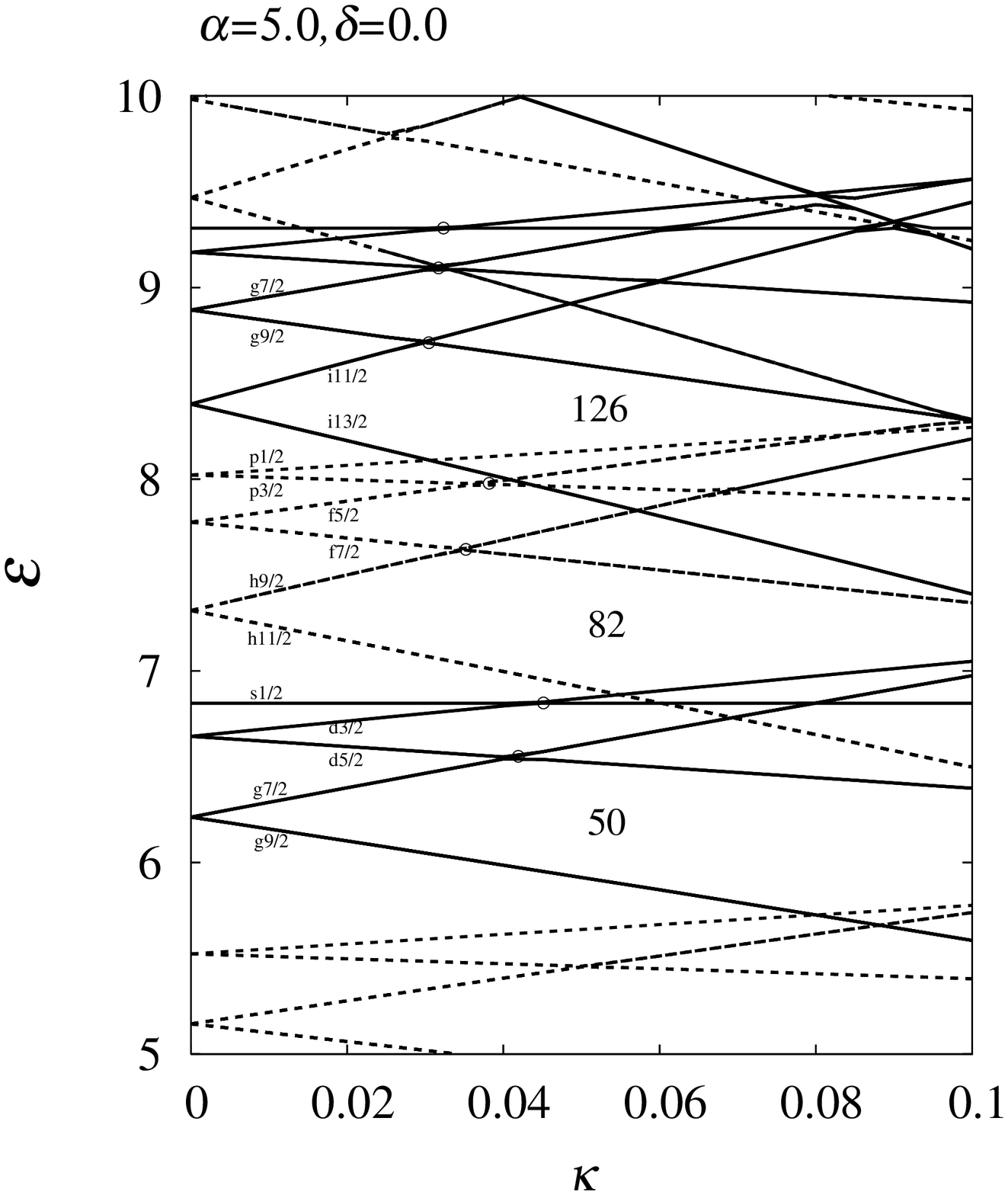}
\end{center}
\caption{\label{fig:pseudo}
Single-particle level diagram against spin-orbit parameter $\kappa$.
Solid and broken curves represent even and odd parity levels,
respectively.  Open circles indicate the crossing points of pseudo
spin-orbit partners.}
\end{figure}

\begin{figure*}[t]
\begin{center}
\includegraphics[width=.55\textwidth]{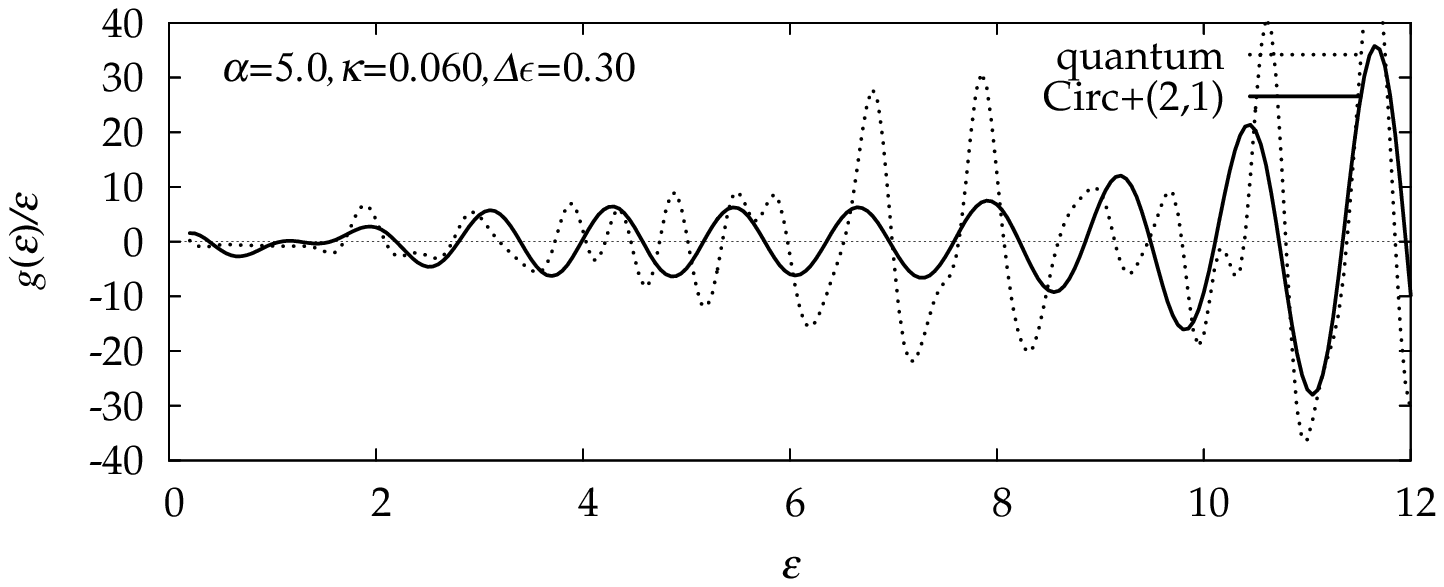}
\includegraphics[width=.15\textwidth]{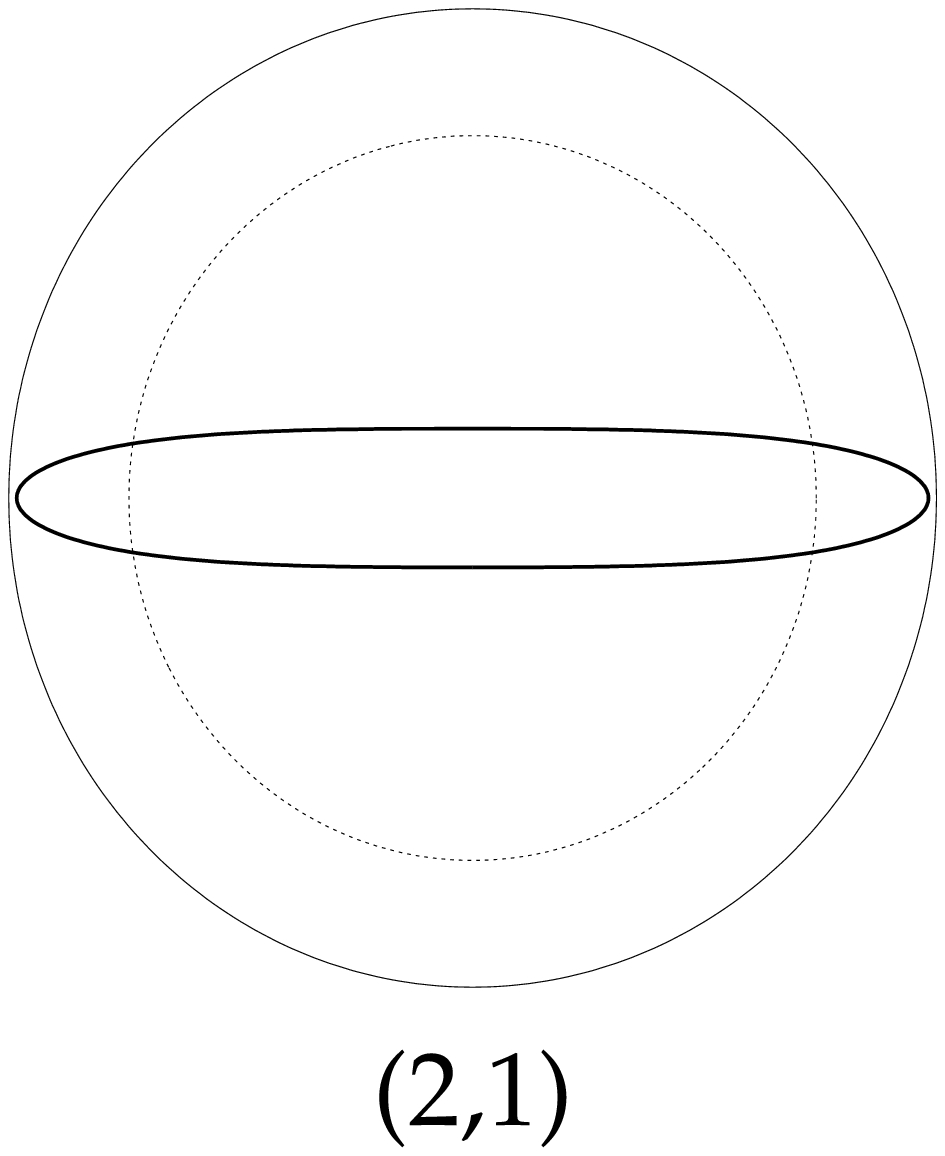} \\
\includegraphics[width=.55\textwidth]{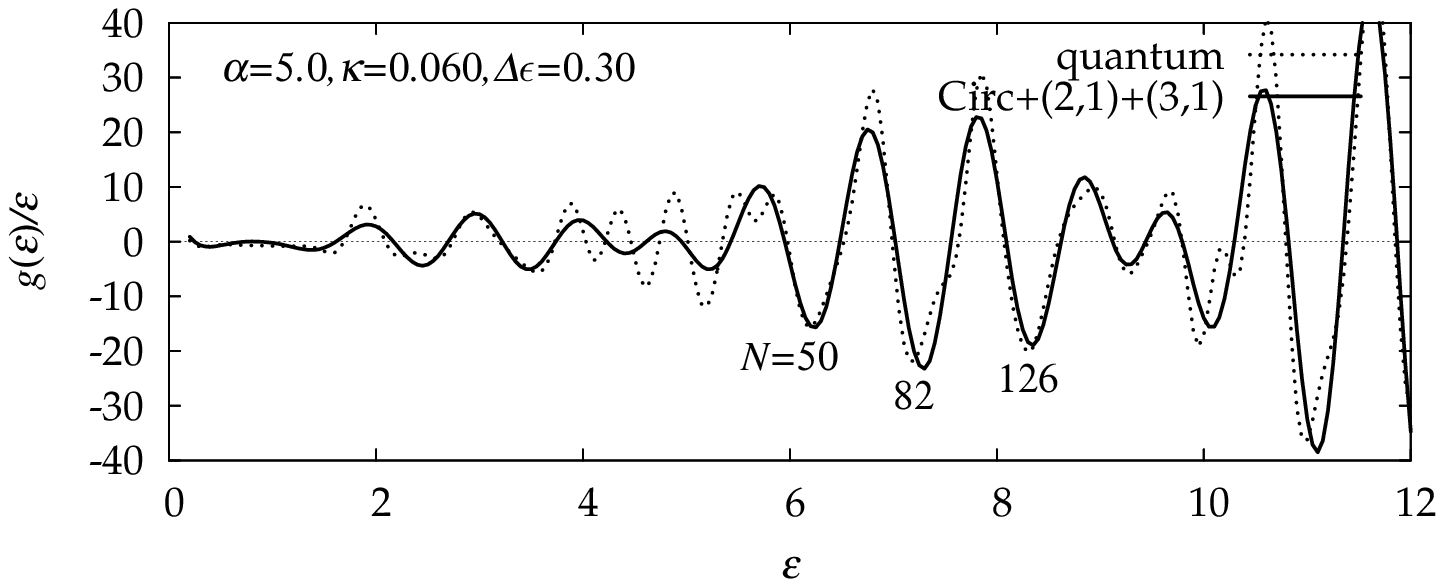}
\includegraphics[width=.15\textwidth]{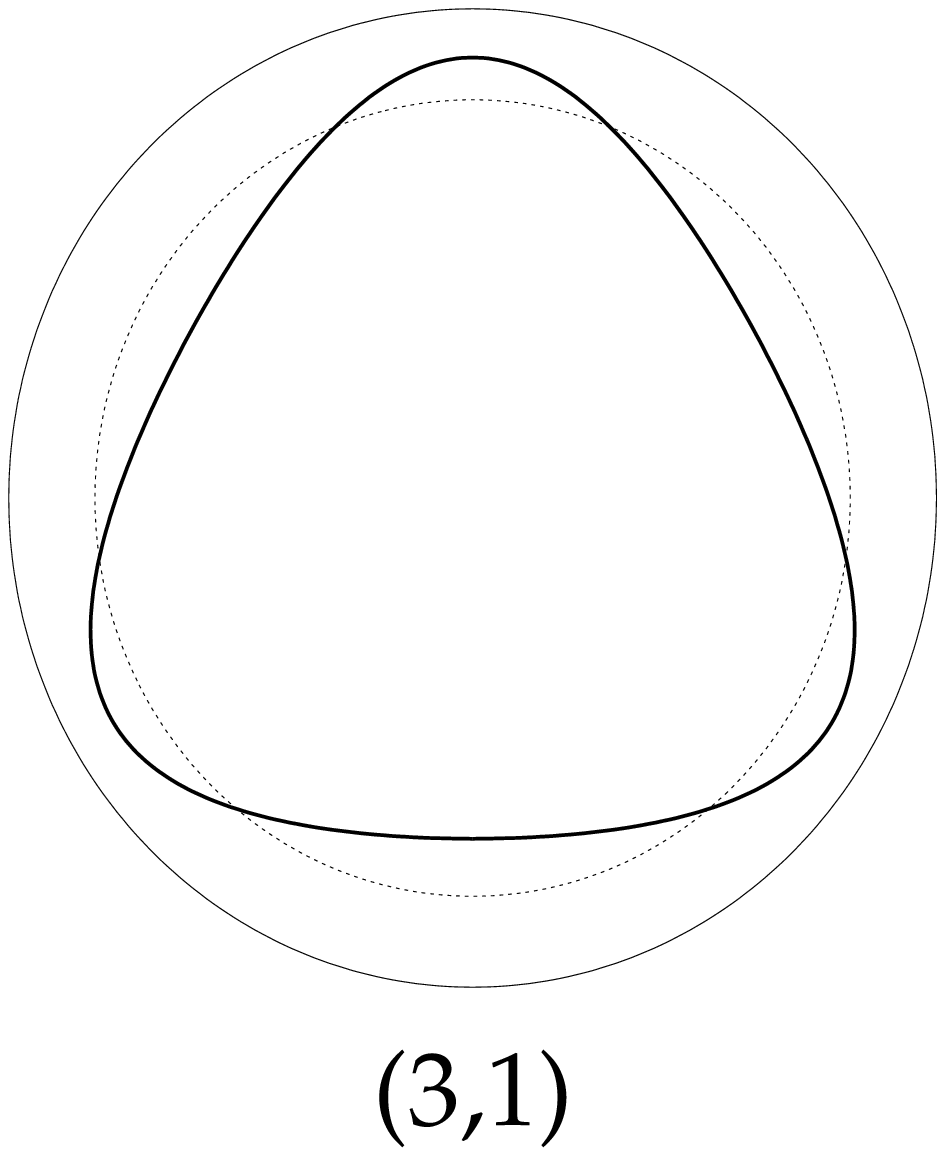}
\end{center}
\caption{\label{fig:sldfit}
Oscillating part of coarse-grained quantum level density is compared
with semiclassical trace formula.  In the top panel, the shortest
circular orbit and oval orbit (2,1) are considered in the periodic
orbit sum.  In the bottom panel, longer circular orbit and triangular
orbit (3,1) are also taken into account.  (In this plot, amplitude
$A_\beta$ and Maslov phase $\nu_\beta$ in trace formula
(\ref{eq:traceformula}) are determined so that they best fit the
quantum result.)  The right panels display the classical periodic
orbits, where the outermost circle represents the boundary of
classically accessible region.}
\end{figure*}

For description of nuclear shell structure, it is very important to
incorporate strong spin-orbit coupling in the mean field.  One can
also apply periodic orbit theory to a system with spin-orbit coupling,
if we utilize SU(2) coherent state path integral
method\cite{Klauder,SK,PZ}.  It introduces classical variable
corresponding to a spin degree of freedom.  Let us write spin vector
in polar coordinate as $\bm{s}=s(\sin\vartheta\cos\varphi,
\sin\vartheta\sin\varphi, \cos\vartheta)$, with $s=\hbar/2$.  One can
choose spin canonical variable as $(q_s,p_s)=(\varphi,s\cos\vartheta)$
and obtains the Poisson bracket relations $\{s_i,s_j\}_{\mathrm{P.B.}}
=\epsilon_{ijk}s_k$, which exactly correspond to the commutation
relation of quantum spin operators.  The Hamiltonian with spin-orbit
coupling is written as
\begin{equation}
{H=\frac{p^2}{2M}+U\left(\frac{r}{Rf(\theta)}\right)^\alpha
-\frac{2\kappa R^2}{\hbar^2}
(\bm{\nabla}V_{so}(\bm{r})\times\bm{p})\cdot\bm{s}}.
\label{eq:hamiltonian_ls}
\end{equation}
In order to make use of scaling property of the central potential, we
adopt spin-orbit potential
\begin{equation}
V_{\rm so}(\bm{r})=U\left(\frac{r}{Rf(\theta)}\right)^{\alpha'}
\end{equation}
and choose the diffuseness parameter $\alpha'=\alpha/2+1$.  With this
choice of spin-orbit coupling, the classical EOM are invariant under
transformation (\ref{eq:scale}) for some special classes of periodic
orbits whose spin does not change its orientation; $\dot{\bm{s}}=0$.
For axially symmetric or reflection-symmetric shapes, such orbits
exist as exact solutions of the EOM.  Using the Hamiltonian
(\ref{eq:hamiltonian_ls}) with suitable $\alpha$ and $\kappa$, we can
reproduce realistic shell structures both for spherical and deformed
nuclei.

\section{Numerical results}

In Fig.~\ref{fig:fourier}, we compare
%examine classical-quantum correspondence by comparing
scaled period $\tau_\beta$'s of classical periodic orbits with Fourier
transform of quantum level density.  Absolute value of Fourier
transform $|F(\tau)|$ are displayed by gray-level image and it
exhibits peaks at dark regions.  We found that all important peaks of
Fourier amplitude are located at scaled periods $\tau_\beta$ of
classical periodic orbits, displayed by curves, and they are
significantly enhanced around the bifurcation points.  Especially, the
Fourier amplitude corresponding to circular and (3,1) orbits are
strongly enhanced around the bifurcation point $\kappa\simeq 0.05$.
It implies a significant growth of gross shell structure around
$\kappa\approx 0.05$.

Figure~\ref{fig:pseudo} shows the single-particle level diagram
against spin-orbit parameter $\kappa$.  As we expected from the
Fourier analysis, remarkable shell structure emerge at about
$\kappa\simeq 0.05$, with large energy gaps corresponding to well
known nuclear magic numbers 50, 82 and 126.  One should also note that
pseudo spin-orbit partners degenerate at almost same values of
$\kappa$.  These degeneracies are responsible for the occurrence of
the above mentioned large energy gaps.  This implies that pseudospin
symmetry\cite{PseudoSpin} is approximately restored in this region.
As we stated in \S1, local restoration of dynamical symmetry is
usually accompanied by bifurcations of classical periodic orbits.
Therefore, we can expect that the pseudospin symmetry might be
connected with periodic orbit bifurcation.

Pseudospin symmetry arise for a certain combination of spin-orbit and
orbit-orbit force parameters in Nilsson Hamiltonian.  In the Nilsson
model, orbit-orbit force is introduced to describe sharp surface of
potential well and therefore connected with surface diffuseness.  In
$r^\alpha$ potential model, bifurcations of periodic orbits occur for
certain combinations of diffuseness parameter $\alpha$ and spin-orbit
coupling parameter $\kappa$.  Therefore, it should be quite natural to
consider that bifurcation of (3,1) orbit which occurs at $\kappa\simeq
0.05$ is a semiclassical origin of gross shell structure
related with pseudospin symmetry.

In order to see the importance of (3,1) orbit for gross shell
structure, we analyze the coarse-grained quantum level density.  In
the left panels of Fig.~\ref{fig:sldfit}, oscillating part of quantum
level density coarse-grained with a smoothing width $\varDelta\cE=0.3$
is plotted with dotted curve.  In the periodic orbit sum, contribution
of long orbits are suppressed by coarse-graining.  For
$\varDelta\cE=0.3$, orbits with scaled periods $\tau_\beta >
2\pi/\varDelta\cE \simeq 20$ are sufficiently suppressed.  The solid
curve in the upper panel of Fig.~\ref{fig:sldfit} represents the
semiclassical level density where only shortest circular orbit (whose
spin is anti-parallel to the orbital angular momentum) and oval orbit
(2,1) (see the right upper part of Fig.~\ref{fig:sldfit}) are taken
into account.  It apparently fail to reproduce the gross pattern of
quantum result.  Next we take into account the contribution of longer
circular orbit (whose spin is parallel to the orbital angular
momentum) and triangular orbit (3,1) (see the right lower part of
Fig.~\ref{fig:sldfit}).  Then we can nicely reproduce the gross
structure of quantum level density as shown in the lower panel of
Fig.~\ref{fig:sldfit}.  Thus we see that this bifurcating orbit
actually play essential role in characterizing the gross shell
structure.

\begin{figure}[t]
\begin{center}
\includegraphics[width=\columnwidth]{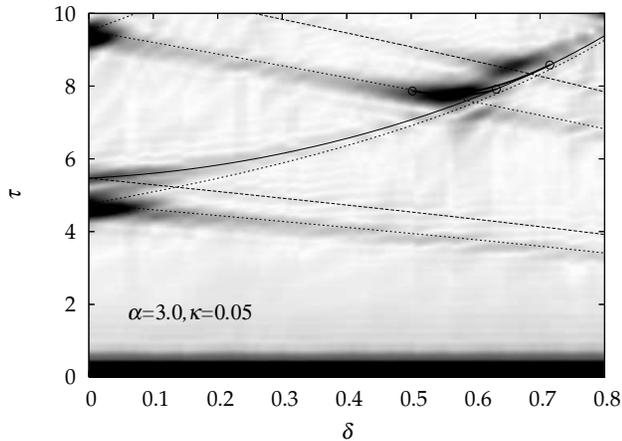}
\end{center}
\caption{\label{fig:ftlmap_sd}
Scaled period $\tau_\beta$ of classical periodic orbits plotted as
functions of deformation parameter $\delta$.  Open circles represent the
bifurcation points.  In the background, Fourier transform of level
density $|F(\tau,\delta)|$ is plotted by gray-scale image.  The
Fourier amplitude exhibits peaks at dark regions.}
\end{figure}

\begin{figure}[t]
\begin{center}
\includegraphics[width=\columnwidth]{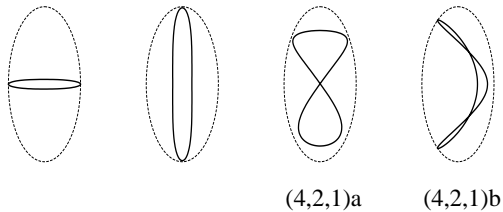}
\end{center}
\caption{\label{fig:orbits_sd}
Periodic orbits responsible for superdeformed shell structure in
$r^\alpha$ model with $\alpha=3.0$ and $\kappa=0.05$.}
\end{figure}

Using the $r^\alpha$ model with spin-orbit coupling, we also analyze
the semiclassical origin of superdeformed shell structure.
Figure~\ref{fig:ftlmap_sd} shows the Fourier transform of level
density as function of deformation parameter
$\delta=3(\eta-1)/(2\eta+1)$, connected with axis ratio
$\eta=R_z/R_\perp$.  We found significant enhancement of Fourier
amplitude at $\delta=0.5\sim 0.6$, corresponding to the bifurcations
of periodic orbits shown in Figure~\ref{fig:orbits_sd}.  With
increasing deformation, oval orbit lying around the short diameter
occur period-doubling bifurcation and new orbits (4,2,1)a and (4,2,1)b
emerge at $\delta\simeq 0.5$, and then they submerge into the oval
orbit lying around the long diameter at $\delta>0.6$.  This situation
is quite similar to the case of the model without spin-orbit
couplings\cite{ASM}.

\section{Summary}

We applied periodic orbit theory to a realistic nuclear mean field
potential model with spin-orbit coupling.  The scaling property of
$r^\alpha$ potential model enables us to investigate quantum-classical
correspondence using Fourier transformation technique.  We found nice
quantum-classical correspondence, and significant roles of
periodic-orbit bifurcations are clarified, also for our systems with
spin-orbit coupling.  Especially, (3,1) orbit bifurcation which occur
for a certain combination of diffuseness and spin-orbit parameters can
be regarded as the semiclassical origin of gross shell structure
related with pseudospin symmetry.  Superdeformed shell structures are
also examined and we found significant roles of bifurcations of periodic
orbits which are similar to those in the model with no spin-orbit
couplings.  It implies that the semiclassical origin of superdeformed
shell structure is essentially accounted for by models without
spin-orbit couplings.


\begin{thebibliography}{99}
\bibitem{Gutzwiller}
M.C. Gutzwiller,
J. Math. Phys. {\bf 8} (1967), 1979; {\bf 12} (1971), 343.
\bibitem{Brack}
M. Brack and R.K. Bhaduri,
{\it ``Semiclassical Physics''}, (Addison-Wesley Reading, 1997).
\bibitem{ExtendedGT}
S.C. Creagh and R.G. Littlejohn,
Phys. Rev. {\bf A44} (1991), 836.
\bibitem{AM}
K. Arita and K. Matsuyanagi,
Nucl. Phys. {\bf A592} (1995), 9.
\bibitem{ASM}
K. Arita, A. Sugita and K. Matsuyanagi,
Prog. Theor. Phys. {\bf 100} (1998), 1223.
\bibitem{Ozorio}
A.M. Ozorio de Almeida and J.H. Hanney,
J. of Phys. {\bf A20} (1987), 5873. \\
\bibitem{Sieber}
M. Sieber,
J. of Phys. {\bf A29} (1996), 4715.
\bibitem{Magner}
A.G. Magner et al, Prog. Theor. Phys. {\bf 102} (1999), 551;
{\bf 108} (2002), 853.
\bibitem{Arita2004}
K. Arita,
Int. J. of Mod. Phys. {\bf E13} (2004), 191.
\bibitem{Klauder}
J. R. Klaudar,
Phys. Rev. {\bf 19} (1979), 2349.
\bibitem{SK}
T. Suzuki and H. Kuratsuji,
J. Math. Phys. {\bf 21} (1980), 472.
\bibitem{PZ}
M. Pletyukhov and O. Zaitsev,
J. Phys. {\bf A36} (2003). 5181.
\bibitem{PseudoSpin}
C. Bhari, J.P. Draayer and S.A. Moszkowski,
Phys. Rev. Lett. {\bf 68} (1992), 2133.
\end{thebibliography}
\end{document}